\documentclass[conference]{ISPASS-Template/IEEEtran}
\usepackage{cite}
\usepackage{amsmath,amssymb,amsfonts}
\usepackage{algorithmic}
\usepackage{graphicx}
\usepackage{textcomp}
\usepackage{xcolor}
\usepackage[hyphens]{url}
\usepackage{fancyhdr}
\usepackage{hyperref}
\usepackage{subcaption}
\usepackage{changes}
\usepackage{listings}
\usepackage{hhline}
\usepackage{enumitem}
\usepackage{dblfloatfix}
\usepackage{makecell}
\newcommand\NOTES{}

\ifdefined\NOTES
\newcommand{\TODO}[1]{\textcolor{red}{\textsf{TODO-#1}}}
\newcommand{\MOE}[1]{\textcolor{orange}{\textsf{MOE-#1}}}
\newcommand{\DANIEL}[1]{\textcolor{magenta}{\textsf{DANIEL-#1}}}
\newcommand{\MUNTAKA}[1]{\textcolor{teal}{\textsf{MUNTAKA-#1}}}
\fi
\unless\ifdefined\NOTES
\newcommand{\TODO}[1]{}
\newcommand{\MOE}[1]{}
\newcommand{\DANIEL}[1]{}
\newcommand{\MUNTAKA}[1]{}
\fi

\newcommand{\sysname}{{eBeeMetrics~}}
\newcommand{\sysnamesec}{{\sc eBeeMetrics}}
\newcommand{\sysnamens}{{eBeeMetrics}}

\newcommand*\circled[1]{\tikz[baseline=(char.base)]{
           \node[shape=circle,fill=lightgray,text=black,draw,inner sep=1pt] (char) {\textbf{#1}};}}

\definecolor{commentgreen}{RGB}{2,112,10}
\definecolor{functioncall}{RGB}{51,34,136}
\definecolor{Black}{rgb}{0,0,0}
\definecolor{NavyBlue}{rgb}{0,0,128}

\title{\sysnamens: An eBPF-based Library Framework for Feedback-free Observability of QoS Metrics}
\IEEEoverridecommandlockouts
\author{
    \IEEEauthorblockN{Muntaka Ibnath\IEEEauthorrefmark{1}, Mohammadreza Rezvani\IEEEauthorrefmark{1}}
    \IEEEauthorblockA{Department of Computer Science and Engineering\\University of California, Riverside\\ Riverside, CA, USA\\ mibna001@ucr.edu, mrezv002@ucr.edu}
    \and
    \IEEEauthorblockN{Daniel Wong}
    \IEEEauthorblockA{Department of Electrical and Computer Engineering\\University of California, Riverside\\ Riverside, CA, USA\\danwong@ucr.edu}
    \thanks{\IEEEauthorrefmark{1}Both authors made equal contributions to this work.}
}

\begin{document}
\maketitle
%%%%%%%%%%%%%%%%%%%%%%%%%%%%%%%%%%%%%%%%
%%%%%%%% -- PAPER CONTENT STARTS -- %%%%%%%%%
\begin{abstract}
Many system management runtimes (SMRs), such as resource management and power management techniques, rely on \textit{quality-of-service (QoS) metrics}, such as tail latency or throughput, as feedback. These QoS metrics are generally neither observable with hardware performance counters nor directly observable within the OS kernel.
This introduces complexity and overhead in instrumenting the application and integrating QoS performance metric feedback with many management runtimes.  

To bridge this gap, we introduced \sysnamens, an eBPF-based library framework to accurately observe application-level metrics derived from only eBPF-observable events, such as system calls. \sysname can be used as a drop-in replacement to decouple system management runtimes from QoS metric feedback reporting, or can supplement existing QoS metrics to better identify server-side dynamics. \sysname achieves a strong correlation with real-world measured throughput and latency metrics across various latency-sensitive workloads. The \sysname tool is open-source; the source code is available at: \url{https://github.com/Ibnathism/eBeeMetrics}.
% We plan to open-source the \sysname tool upon publication of this work. \MUNTAKA{Edit this?}
%Overall, \sysname provides a non-intrusive approach to gain insights into application performance for system management.  
\end{abstract}
\section{Introduction}
\label{section:Introduction}
Modern data centers require significant amounts of profiling, tracing, measurement, and observability to maintain performance, reliability, and efficiency. These telemetries are then used to guide \textit{system management runtime (SMR)} frameworks for resource management, scheduling, power management, etc. Application performance is directly influenced by these runtimes. For instance, by adjusting core frequency or reallocating cores, we can trade off between performance, dynamic power consumption, and utilization~\cite{8675218,10.1145/2830772.2830797,9251983,10.1145/3297858.3304005,9065583,9499723}.

%\DANIEL{Define application-level metrics a bit first.}
Many existing system management runtime frameworks require quality-of-service (QoS) feedback to guide management decisions, such as tail latency or throughput. 
%Many of these works require feedback metrics directly from the application or the client. 
This can be problematic as applications may need to be instrumented to report QoS metrics, and client-provided QoS metrics may incur overheads. Furthermore, depending on where the system management runtime operates (userspace or kernel space), it may be intrusive to provide QoS metrics. For example, we cannot readily pass feedback metrics directly to in-kernel dynamic power management drivers or Linux schedulers. This would require passing through \texttt{sysfs}, which incurs frequent and significant syscall overheads that make timely reporting of feedback metrics unfeasible. 

% \noindent\textbf{Challenges: } 
%Existing QoS feedback mechanisms to system management runtimes can be intrusive or impractical to convey. For example, QoS-aware DVFS techniques tend to rely on tail latency distributions, which the Linux system cannot directly obtain, leading to much of the work being evaluated through simulation or a limited-feature evaluation in real servers~\cite{10.1145/2830772.2830797,8675218,10.1145/2934583.2934616}. Thus, there is a need to provide low-overhead, non-intrusive mechanisms to provide QoS metrics to system management runtimes.

Recently, eBPF~\cite{eBPF-Website} has emerged as a major kernel technology framework used to build many tools for observability, monitoring, security, etc. eBPF (extended Berkeley Packet Filter) is a Linux subsystem that enables custom user-space programs to run within a sandboxed in-kernel virtual machine. Specifically, \textit{eBPF programs} are event-driven and are triggered by certain \textit{system events}, such as network events, system calls, function entry/exit, etc. Furthermore, eBPF programs are extremely lightweight. For example, Facebook actively runs 40 eBPF programs per server, with 100s of additional on-demand eBPF programs, and Netflix runs 14 active eBPF programs per server instance~\cite{Starovoitov}. By being able to safely extend the Linux kernel, eBPF has been the foundation of numerous frameworks that aim to provide unobtrusive observability, security, monitoring, and tracing functionality for many distributed systems, such as microservices and cloud-native systems~\cite{Facebookincubator, Falcosecurity, githubGitHubApacheskywalkingrover, githubGitHubPixieiopixie}.

Prior work~\cite{Paper:kernelobservability} has shown that from eBPF-derived \textit{traces of syscalls} and offline analysis, it is possible to accurately estimate a proxy throughput metric (requests per second) that is highly correlated with the actual RPS, demonstrating the potential for eBPF observability into application-level metrics. However, several limitations of this work limit its real-world usefulness. First, the estimated RPS is not 1:1 with the real RPS; rather, it serves as a well-correlated proxy. Second, only throughput-based metrics can be estimated, and not latency-based metrics, which are crucial for resource management of latency-sensitive workloads. Third, the proxy metrics were derived offline through eBPF traces of syscalls, which limits utility since system management runtimes require real-time streaming QoS feedback.

\begin{figure*}[!t]
\centering
%\vspace{-5mm}
\includegraphics[width=0.65\linewidth]{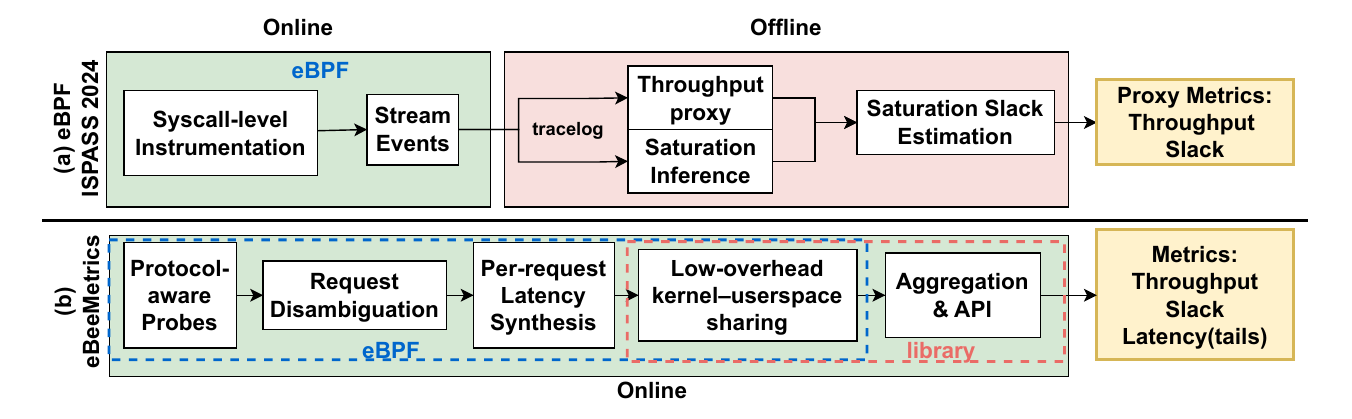}
\vspace{-3mm}
\caption{(a) Prior work~\cite{Paper:kernelobservability} demonstrates observability of throughput-based proxy metrics through offline analysis of eBPF-provided traces. (b) \sysname enables a fully online solution for observability of both throughput and latency metrics.}
\vspace{-5mm}
\label{fig:ispass-vs-eBee}
\end{figure*}

In this work, we overcome these limitations with \sysnamens, a fully online eBPF-based library solution that transparently analyzes real-time streaming eBPF events of target workloads to accurately estimate throughput and latency metrics. % enables accurate estimation of both throughput and latency application-level metrics. Our tool transparently analyzes real-time streaming eBPF events of target workloads to accurately estimate throughput and latency metrics.   
%\DANIEL{TODO: Revisit and update later}
The key contributions of this paper are: 
\begin{itemize}[leftmargin=12pt,noitemsep,topsep=0pt]
\item We present \sysnamens, an online eBPF-based library that enables observability into QoS metrics (such as tail latency, throughput, etc.) without any application instrumentation or direct feedback. \sysname presents an API interface for system management runtimes to obtain QoS telemetry without direct reporting from applications. 
\item The key to enabling accurate observability of application QoS metrics is in disambiguating request boundaries and identifiers from streaming eBPF events. 
% This is achieved through careful analysis of the most common client-server protocols, such as HTTP1.1/REST and gRPC. 
In this work, we target HTTP/1.1/REST and gRPC/HTTP/2, and more broadly assume that a workload exposes either reliable syscall-level request lifecycles or stable user-space request start/end hooks that can be probed with eBPF.
% \item We tackle the inherent challenges of tracing in gRPC, where traditional syscall-based approaches fall short due to connection reuse and binary framing, unlike the more traceable patterns in HTTP/1.1.
% \item We demonstrate how request-level throughput and latency metrics can be derived under varying workloads using real application inputs.
\item We demonstrate the efficacy of \sysname across a range of real-world latency-critical workloads. \sysname can accurately obtain QoS metrics with minimal overhead, effectively decoupling QoS feedback from the applications. 
\end{itemize}

% 
% Latency metrics (prior work: throughput)
% Library (prior work: traces)
% Demonstrate integration into PARTIES

%\DANIEL{Revisit and update}
%The remainder of the paper is organized as follows: Section~\ref{section:Background} presents the background and related work. Section~\ref{section:Network} introduces our core insight towards identifying request boundaries from eBPF observable events. Section~\ref{section:Framework} demonstrates \sysname and details its implementation. Section~\ref{section:Evaluation} evaluates the effectiveness of \sysnamens, while Section~\ref{section:Case Study} discusses its integration with an exemplar system management runtime. Section~\ref{section:Conclusion} concludes the paper. 
\section{Background and Motivation}
\label{section:Background}

\subsection{QoS Metric Challenges for System Management Runtimes}
Many system management frameworks (such as resource management, power management, or performance debugging) need to be aware of quality-of-service (QoS). Thus, these resource management frameworks often require feedback of QoS metrics (such as latency or throughput) in order to guide management decisions. This often requires client-provided metrics, or application instrumentation, which limits the practicality of system management runtimes~\cite{9499723, 10.1145/2830772.2830797, 10.1145/2749469.2749475, 7920819, 6307780, 16682, 10.1145/2644865.2541944,8425223, 10.1145/2485922.2485974, 10.1145/2556583, 9355316, 7551416, 10.1145/1508244.1508269, 6983037, 10.1145/3046682, 1598114, 10.1145/2934583.2934616, 6853237, 10.1145/2806777.2806848, 196288, zhan2016carb,6493613}.

% Despite the importance of performance, the need to minimize power consumption while meeting requirements is becoming increasingly sophisticated through the use of technologies such as DVFS~\cite{9355316, 10.1145/1508244.1508269, 6983037}.
%From per-core power management~\cite{10.1145/3046682} and multi-core~\cite{1598114} to metric-aware techniques~\cite{10.1145/2934583.2934616, 6853237, 10.1145/2806777.2806848, 196288, zhan2016carb}. 

%We envision that \sysname can provide a drop-in replacement to decouple the need for direct application-level latency/performance feedback.
%\noindent\textbf{Performance modeling:}

For example, in many environments, QoS metrics are not directly available, or a cloud provider doesn't have visibility into client workloads. 
Prior works, such as PACT~\cite{10.1145/3419111.3421274}, assign applications to multiple classes (latency-critical vs. best-effort) and then allocate different resources to those classes. Latency-critical workloads are still satisfied by avoiding reducing the frequency of such workloads and only applying power savings techniques to best-effort workloads, which significantly limits the potential power-saving benefits.

Other studies~\cite{6926798, 6730745, 6212065, Kounev2020, 9251983, 10.1145/3445814.3446693, 9773201} have demonstrated the ability to predict workload types and estimate QoS requirements through workload characterization and performance models. However, these techniques are not generalizable and require significant instrumentation or pre-training of predictive models. Therefore, there has been recent interest in providing better observability into application QoS metrics while avoiding the need for direct QoS feedback from applications. 

%\sysname provides an alternative solution that decouples system management from QoS feedback reporting by estimating application-level metrics using eBPF observable events.  

%\sysname provides a mechanism to provide a generalized estimation of application-level metrics while maintaining applications as a black box.

%\subsection{Linux tracing}
%Modern Linux systems offer a wide range of tracing and observability tools for capturing system behavior and diagnosing performance issues~\cite{Website:juliaevans}. \texttt{strace} is a widely used tool for observing kernel syscalls, but its high overhead from context switches makes it unsuitable for low-latency applications. Alternative tools such as \texttt{perf}~\cite{demelo2010perf} enable performance counter collection and statistical profiling, while \texttt{ftrace}~\cite{rostedt2010ftrace} allows detailed tracing of function calls in the kernel with filtering capabilities. Tools like \texttt{SystemTap} \cite{systemtap2022} and LTTng \cite{lttng2023} have also been proposed to balance trace granularity and system impact, but often require kernel modules or debug symbols, which limit ease of deployment. These limitations have influenced the move toward modern, lower-overhead frameworks like eBPF~\cite{eBPF-Website}.

\subsection{eBPF Overview}
%eBPF provides a promising path towards observability of QoS metrics in a low-overhead, non-intrusive manner. %Prior work~\cite{Paper:kernelobservability} demonstrated that eBPF-generated traces and offline analysis can capture syscall activities that can be used as a proxy for throughput, highlighting the potential for eBPF-derived metrics. 
eBPF enables sandboxed programs to execute within the Linux kernel, allowing developers to extend kernel functionality without modifying kernel source code or loading kernel modules. %More recently, support for eBPF has also been introduced on Windows platforms~\cite{ebpf-for-windows}. 
eBPF is flexible and incurs minimal overhead, as demonstrated by widespread adoption in industry and emerging use cases~\cite{eBPF-Website,9359159, 10.1145/3477132.3483542,10.1145/3544216.3544259,9335808, 9724089, 8416289, 10.1145/3472883.3487014}.
% eBPF programs essentially extend the capabilities of the kernel without requiring source code changes to the kernel. 

eBPF programs are executed when triggered by specific \textit{system events} within the kernel, such as system calls, interrupts, or network events. The programs are attached to pre-defined hooks inside the kernel, and a specific set of instructions is executed when an event associated with the corresponding hook takes place. Furthermore, eBPF programs can be attached at nearly any point within the kernel or user-space applications~\cite{eBPF-Website}. eBPF provides numerous built-in hooks through tools such as BPF Compiler Collection (BCC)~\cite{noauthor_2016-se} and bpftrace~\cite{noauthor_undated-ky} that can be used to monitor different aspects of the system. % For example, hooks exist to monitor system calls, interrupts, TCP/UDP connections, network packet distributions, packet drops, file systems, schedulers, etc. 
eBPF doesn't require any instrumentation in the application code and allows run-time tracing. Low-overhead data sharing mechanisms between kernel space and user space are provided through BPF maps~\cite{gregg2019bpf}, such as ring buffers and perf buffers~\cite{starovoitov2015plumbers}. %\DANIEL{This sentence sounds like BPF map enables tracing. Is this right? Sounds weird...} %\DANIEL{Add some citations to BPF map and perf buffers. What are perf buffers?? Are these interfaces to trace? This last sentence is a bit confusing. \MUNTAKA{Added a bit of details on BPF maps and perf buffers}} 
%\subsubsection{eBPF applications}
%eBPF can be a very efficient tool for monitoring the system behavior and detecting failures~\cite{9335808, 9724089, 8416289}.
%Aside from monitoring applications, eBPF has been used as a lightweight virtual machine inside the kernel to run programs as efficiently as possible~\cite{9359159, 10.1145/3477132.3483542} and to provide light-weight serverless functions~\cite{10.1145/3544216.3544259}.
%By leveraging the benefits of eBPF, \sysname can monitor syscalls while estimating application-level metrics within the eBPF environment.
%\noindent\textbf{System Calls as a tool and a threat:}
%Syscalls can provide a great deal of information regarding the behavior of an application.
%\sysname uses a syscall to extract valuable information for performance feedback; however, they can be used in malicious ways.
%Some studies aimed to reduce the attack surface in the operating system caused by system calls~\cite{Ghavamnia2020TemporalSC, 263868, Ghavamnia2020ConfineAS, DeMarinis2020sysfilterAS}; whereas other studies used modern machine learning approaches to detect anomalous activities ~\cite{Peddoju2020}.
%As well as this, studies have been conducted to enhance the security of system calls by modifying the operating system architecture~\cite{9251949}.}

\begin{figure}[!b]
\centering
\vspace{-6mm}
\includegraphics[width=\linewidth,trim={9mm 4mm 9mm 2mm},clip]{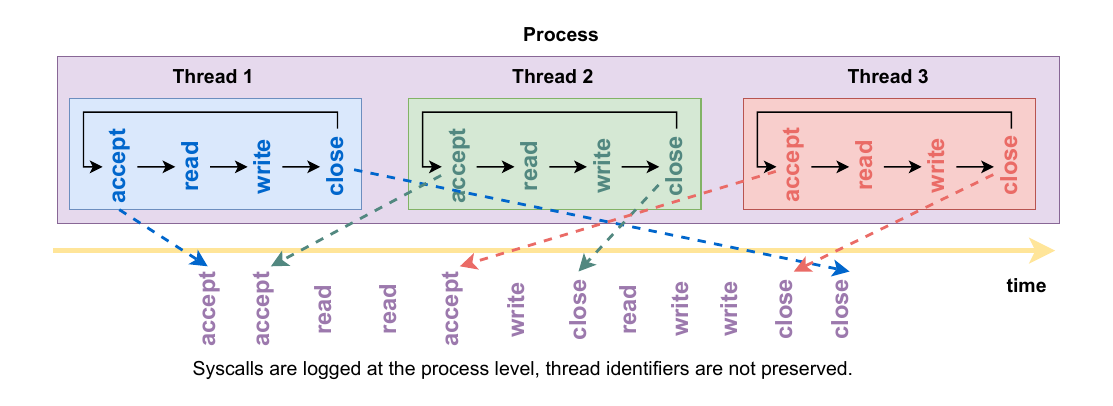}
\caption{Key syscalls from request processing using HTTP/1.1 protocol. Syscalls from different worker threads can be interleaved, obfuscating request boundaries due to syscalls only preserving process association (PID) and not threads (TID). }
%\vspace{-5mm}
\label{fig:http11-mapping}
\end{figure}

\subsection{Challenges toward eBPF for QoS metrics observability}
As shown in Figure~\ref{fig:ispass-vs-eBee}(a), prior work~\cite{Paper:kernelobservability} has shown that from eBPF \textit{traces of syscalls}, it is possible to estimate a proxy throughput metric (requests per second) that is highly correlated with the actual RPSs. From this proxy RPS, saturation, and slack metrics were also derived to demonstrate the potential for eBPF to provide proxy QoS metrics. 

However, these proxy metrics have several limitations that hinder deployment in real system-management runtimes. The inferred RPS is not a one-to-one match with ground-truth throughput and should be treated as a correlated surrogate. In addition, the eBPF-based approach reliably reconstructs throughput-oriented signals but does not recover latency metrics, which are essential for managing latency-sensitive workloads. Finally, prior work derives these proxies via offline syscall-trace analysis, whereas practical control loops require real-time, streaming feedback.

\subsubsection{Obfuscation of request boundary}
The root issue of prior work~\cite{Paper:kernelobservability}, which leads to these limitations, is its inability to identify request boundaries. Figure~\ref{fig:http11-mapping} illustrates this issue. Latency-critical workloads are request-response services that process an incoming request and return a response. Multiple worker threads can handle these requests. Figure~\ref{fig:http11-mapping} illustrates a server process with three worker threads. Throughout the course of processing these requests, the worker threads would trigger various syscalls\footnote{Further details regarding client-server protocols and system call event triggers are discussed in Section~\ref{section:Network}.}. For example, each thread can open and close a connection (with \texttt{accept} and \texttt{close} syscalls), read the request payload (with \texttt{read} syscall), and then write the response payload (with \texttt{write} syscall). 

For a single thread, the beginning and end boundary of a request can be identified by when it enters the server (\texttt{accept}) and when it exits the server (\texttt{close}). However, worker threads operate concurrently, and the syscalls ordering between threads may interleave; thus, the observed order of \texttt{accept} syscalls may not correspond with the \texttt{close} syscall, as highlighted in the figure. Furthermore, syscalls only contain metadata to identify the associated \textit{process} (PID) and not the specific \textit{thread} (TID). Thus, the request boundaries can not be readily identified from existing syscall traces, making the observability of latency metrics difficult. Although these syscalls no longer correspond, prior work~\cite{Paper:kernelobservability} found that inter-syscall times of \texttt{send}, \texttt{recv}, and \texttt{epoll\_wait} syscalls can be a good signal for estimating a proxy throughput metric.

\textbf{\textit{Therefore, the key to unlocking latency QoS metric observability is to efficiently disambiguate the requests using only eBPF observable events and metadata.}} 
Once requests can be disambiguated, it provides a foundation to build out a fully online eBPF library for QoS metric observability.
Figure~\ref{fig:ispass-vs-eBee}(b) illustrates how this work (\sysnamens) improves upon prior work~\cite{Paper:kernelobservability} to fully enable eBPF observability into application QoS metrics.
Besides request disambiguation, we also require generalizability to different client-server protocols and the ability to efficiently extract and calculate various QoS metrics online from streaming eBPF events. 

%The core challenge here in leveraging eBPF for these metric collection lies in identifying the appropriate probes to observe request-level behavior. eBPF supports both kernel-space and user-space tracing through \textit{kprobes} and \textit{uprobes}, respectively; therefore, selecting the correct probe points is critical for accurate and efficient instrumentation. Moreover, eBPF programs execute within a constrained, sandboxed virtual machine, which imposes limitations such as restricted data structures and a lack of floating-point operation support. As a result, eBPF-based solutions must maintain low overhead and operate within these constraints while still delivering a comprehensive and user-friendly library interface for performance monitoring.

\begin{figure}[!t]
\centering
\includegraphics[width=0.9\linewidth]{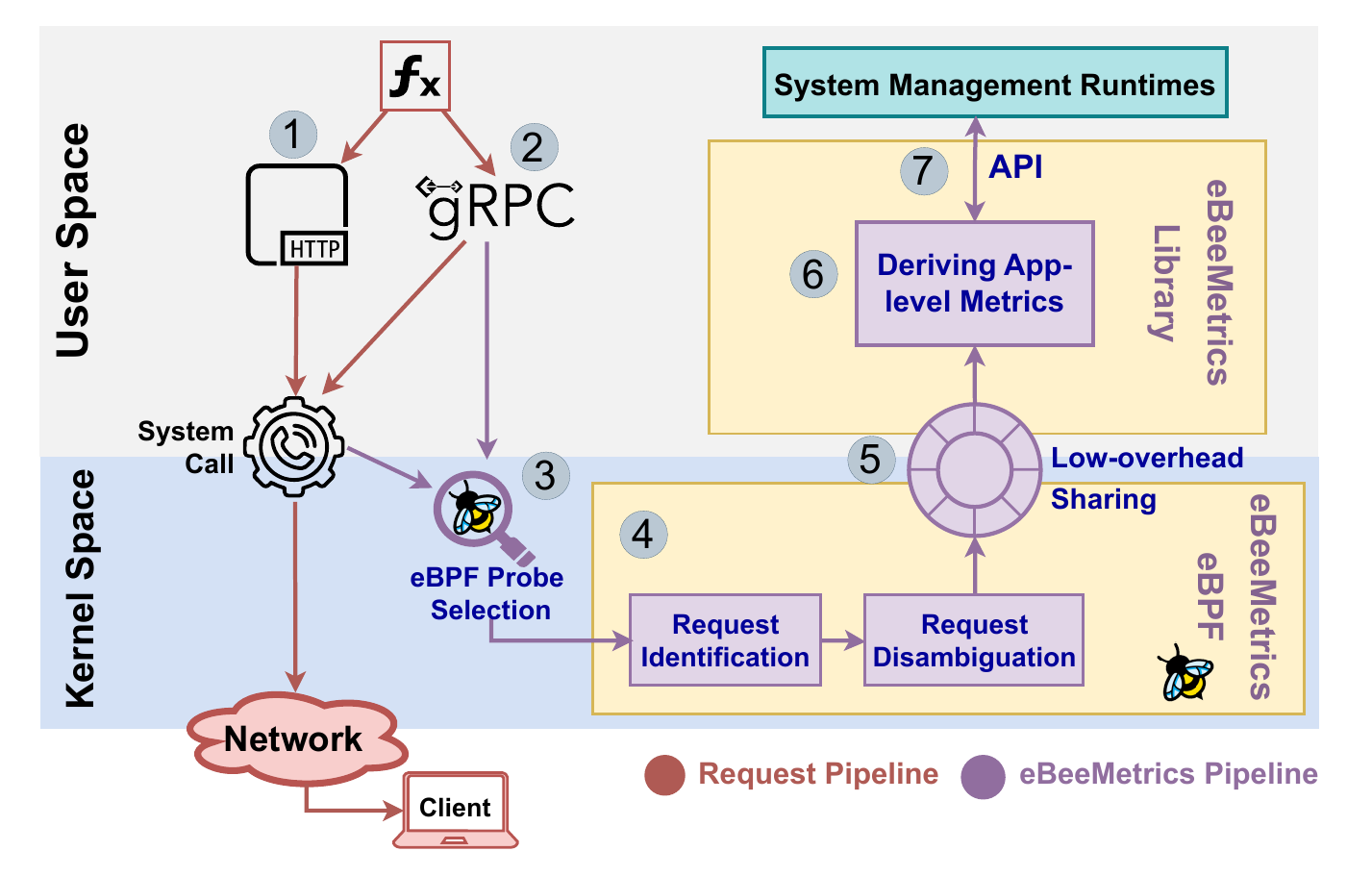}
\vspace{-3mm}
\caption{Detailed overview of \sysnamens. }%\sysname consists of two main components: (1) the eBPF program, which identifies individual request-level latencies and (2) the library, which computes application-level metrics (such as tail latency, throughput, etc.) and interfaces with system management runtimes.}
\vspace{-5mm}
\label{fig:overview}
\end{figure}

\section{\sysnamesec}
\label{section:Framework}

In this section, we present \sysnamens, an eBPF-based library for observability of QoS metrics for latency-sensitive applications. %\sysname can accurately provide request-level metrics (such as latency and throughput) to system management runtimes without direct feedback from applications, enabling responsive and direct QoS-aware management policies, even in the Linux kernel. 
%These applications may employ various network protocols and configurations, leading to diverse and often unpredictable runtime behavior. Such variability presents significant challenges in accurately collecting the performance metrics required by system management runtimes. 
As highlighted previously, effective metric collection relies on the ability to isolate individual requests, correlating them with their corresponding responses, and thus enabling observability into per-request latency. %\sysname addresses these challenges by enabling precise per-request latency measurement, offering detailed visibility into application behavior without necessitating any modifications to the application code or kernel. 
\sysname achieves this by identifying pertinent system events in how request-response protocols handle requests. 
The \sysname library exposes a lightweight interface for system management runtimes to obtain accurate, real-time feedback metrics with negligible overhead to the application.

%\subsection{Overview}

Figure~\ref{fig:overview} illustrates the detailed design of \sysnamens, which is designed to handle diverse application behaviors across different client-server protocols. %To provide a deeper understanding of \sysname, Figure~\ref{fig:overview} outlines its end-to-end architecture. 
When an application handles a client request, it invokes a protocol-specific library function, either HTTP (REST) or gRPC, illustrated in \circled{1} and \circled{2}. Regardless of the protocol, this triggers system calls that process the request and ultimately facilitate network communication with the client. \sysname leverages eBPF probes placed either in kernel space (via kprobes) or user space (via uprobes) to extract request identifiers. These identifiers are then processed (\circled{4}) to correlate timestamps and determine request boundaries in real time. The resulting per-request measurements are stored in an asynchronous eBPF ring buffer in shared memory (\circled{5}), enabling efficient communication between the kernel and user space. The \sysname library in user space (\circled{6}) continuously polls, aggregates, and updates recent request statistics while preserving request history. It also exposes lightweight APIs (\circled{7}), allowing system management runtimes to retrieve accurate performance metrics on demand, with minimal overhead.

\subsection{Identifying individual requests from eBPF events}
% \subsection{Request Disambiguation and Identification}
\label{section:Network}
As previously discussed, successfully disambiguating request forms the foundation of observability into application QoS metrics. 
Since each request and response traverses the network, the timing and ordering of network events naturally align with request boundaries. The key challenge is in identifying the necessary identifiers from eBPF observable events and metadata. 
We first describe how eBPF probes expose relevant network activity and then show how we adapt our tracing strategy to common client-server protocols.

%By tapping into these low-level operations, we can infer when a request begins and ends without any modifications to application code. 
%In particular, we leverage eBPF to instrument kernel and user space hooks on network events and capture precise timestamps for each request. 
%This approach lets us infer per-request latency and throughput in real time with minimal overhead to replace application-provided feedback. 
%In the remainder of this section, 

\subsubsection{How do clients and servers communicate?}
\begin{figure}[!t]
\centering
\includegraphics[width=\linewidth]{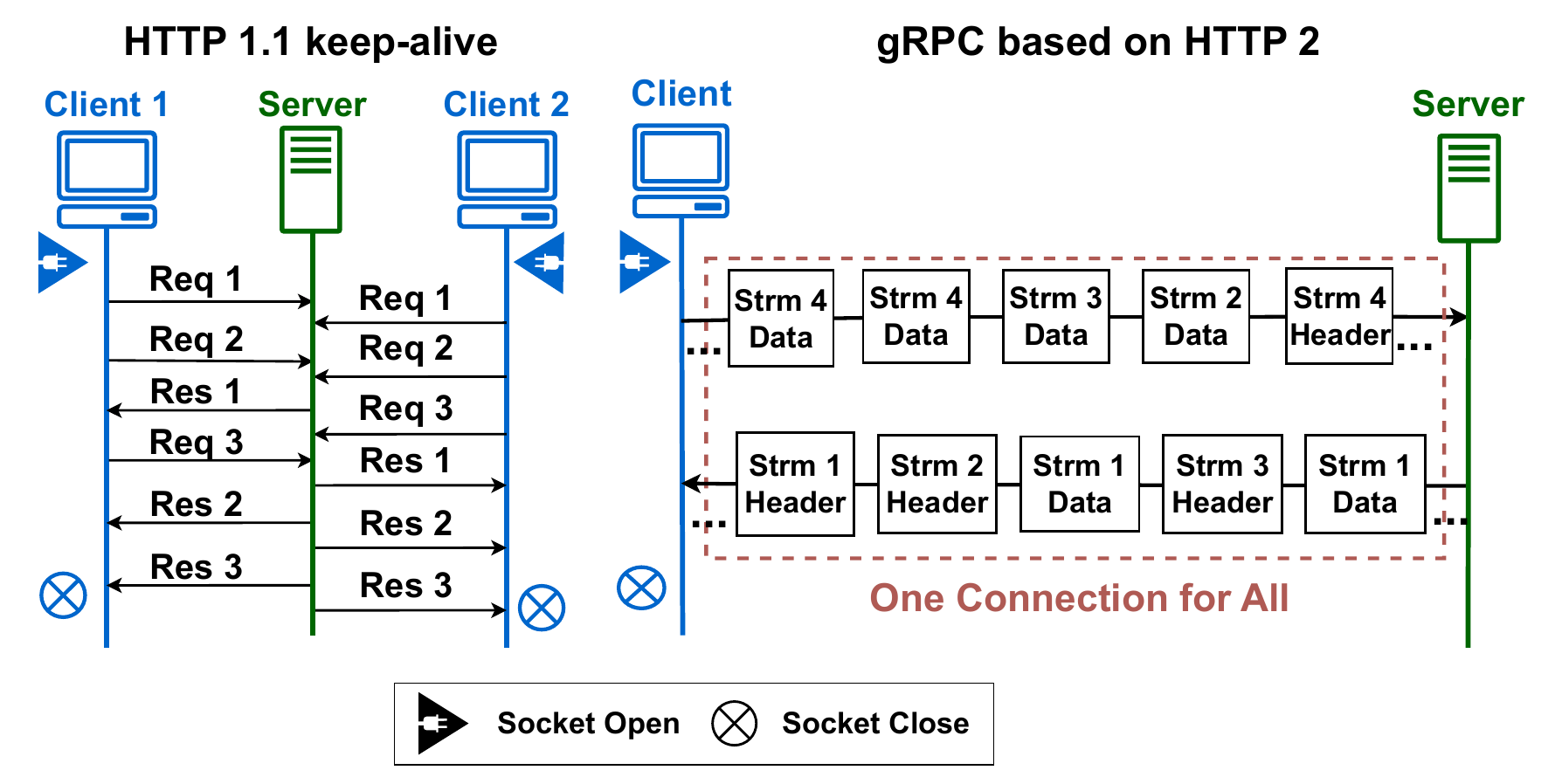}
\caption{Client-Server communication protocols of HTTP/1.1 and HTTP/2.} %Concurrent request handling makes identifying individual requests from eBPF events non-trivial.}%, as events of different requests are interleaved. 
\vspace{-5mm}
\label{fig:http1.1-vs-http2}
\end{figure}

To identify the proper eBPF probes to use (\circled{3}), we first present a background on how clients and servers communicate for common web applications (\circled{1}, \circled{2}). 
Client and servers communicate most commonly through HTTP/REST protocols or remote procedure calls (RPC). Figure~\ref{fig:http1.1-vs-http2} illustrates how these two common protocols are used for request-response, including the socket connections. 
% on both the server and the client side.

\lstdefinestyle{trace}{
  language={},
  basicstyle=\ttfamily\scriptsize,
  columns=fullflexible,
  breaklines=true,
  frame=single,
  keepspaces=true,
  numbers=left,
  numberstyle=\tiny\ttfamily,
  stepnumber=1,
  numbersep=8pt,
  emph={sendto,recvfrom,writev,readv,close,accept4,retval,fd,pid},
  emphstyle=\bfseries
}
\begin{figure*}[!t]
\centering

\begin{minipage}{0.9\textwidth}
\begin{lstlisting}[style=trace]
tritonserver-479363 [000] d...1 13485159.048749: bpf_trace_printk: sendto: pid=478966 fd=10 len=17
tritonserver-479405 [009] d...1 13485159.048910: bpf_trace_printk: recvfrom: pid=478966 fd=9 len=17 buf="\0'\r\0\204\334u\0\0\360N\306i\276W\0\0"
tritonserver-479405 [009] d...1 13485159.048968: bpf_trace_printk: writev: pid=478966 fd=63
tritonserver-479405 [009] d...1 13485159.051795: bpf_trace_printk: close: pid=478966 fd=63
tritonserver-479405 [009] d...1 13485159.051806: bpf_trace_printk: close return: pid=478966 retval=0
tritonserver-479413 [008] d...1 13485159.107110: bpf_trace_printk: accept4: pid=478966 retval=63 buf="\0'\r\0\204\334u\0\0\360N\306i\276W\0\0"
tritonserver-479405 [009] d...1 13485159.171612: bpf_trace_printk: readv: pid=478966 fd=64
tritonserver-479364 [028] d...1 13485159.127098: bpf_trace_printk: close: pid=478966 fd=64
tritonserver-479279 [000] d...1 13485159.171294: bpf_trace_printk: sendto: pid=478966 fd=10 len =17
tritonserver-479405 [009] d...1 13485159.171542: bpf_trace_printk: recvfrom: pid=478966 fd=9 len =17 buf="\0'\r\0\214\334u\0\0\360N\306i\276W\0\0"
tritonserver-479405 [009] d...1 13485159.171612: bpf_trace_printk: writev: pid=478966 fd=63
tritonserver-479405 [009] d...1 13485159.172001: bpf_trace_printk: close: pid=478966 fd=63
tritonserver-479405 [009] d...1 13485159.172009: bpf_trace_printk: close return: pid=478966 retval=0
\end{lstlisting}
\end{minipage}
\vspace{-3mm}
\caption{Example of raw eBPF syscall trace logs from eBPF trace pipe. A key insight of \sysname is in identifying metadata from eBPF observable events that can allow us to tease out individual requests and their timing.}
\label{code:kprobe-traces}
\vspace{-3mm}
\end{figure*}

\paragraph{REST (HTTP/1.1)} Clients and servers that communicate through REST APIs are built on top of the HTTP/1.1 protocol, a stateless application layer protocol that provides a persistent connection\cite{Website:http}. 
Figure \ref{fig:http1.1-vs-http2} (left) illustrates the HTTP/1.1 protocol. When a client wants to connect to a server, it first opens a TCP socket connection to that server. Once it is open, the client can send a request to the server and receive a response. 
HTTP/1.1 can be configured to either create one TCP connection for each request or reuse a single persistent connection for multiple requests via keep-alive, known as pipelining. Under a persistent connection, the client does not need to wait for a response before sending subsequent requests. Once all requests and responses are issued, the client can then close the TCP socket connection.

% \begin{figure*}[!t]
% \centering
% \includegraphics[width=0.75\linewidth]{Files/Attachments/trace_kprobe.png}
% \caption{Example of raw eBPF syscall trace logs from eBPF trace pipe. A key insight of \sysname is in identifying metadata from eBPF observable events that can allow us to tease out individual requests and their timing.\TODO{Mark each line to use as an example, or maybe use code/verbatim}}
% \vspace{-3mm}
% \label{fig:kprobe-traces}
% \end{figure*}

\paragraph{gRPC (HTTP/2)} %HTTP/1.1 suffers from performance limitations due to head-of-line blocking, where responses must be delivered in the same order as the corresponding requests. 
HTTP/2 improves on HTTP/1.1 by using true multiplexing, which refers to multiple request/response streams sharing a single persistent TCP connection\cite{Website:grpc}. gRPC builds on this by enabling many concurrent RPCs over the same connection. After the connection is established, requests and responses are carried in \textit{streams} that may be processed out of order as illustrated in Figure~\ref{fig:http1.1-vs-http2} (right). This boosts efficiency but makes it harder to trace per-request server-side metrics. HTTP/2 also adds header compression and a binary framing format, improving latency and efficiency but making packet-level inspection more challenging. Finally, gRPC implementations vary in how they use connections (e.g., multiple sockets per stream vs. a single reused connection), further complicating accurate metric extraction and motivating a generalizable approach that handles diverse client-server protocol behaviors.

\subsubsection{What to trace? How to disambiguate requests?}
One of the core challenges in isolating request boundaries across diverse protocols lies in selecting the appropriate type of probe. \sysname aims to identify a minimal subset of probes that allows us to overcome the challenge. %By accurately identifying request boundaries, we will be able to estimate the latency of each request. 
In particular, to disambiguate requests, we must identify (1) the \textit{boundary} of a request (i.e., the start and end of request processing on the server) and (2) a unique identifier that allows us to associate the corresponding start and end events for each request.

% \DANIEL{To disambiguate requests, we need to identify (1) the \textit{boundary} of a request (start and end of request processing in the server) and (2) a unique identifier for the request (so we can associate start and end request events).}

\paragraph{REST/HTTP Request \underline{Boundary}}
%For applications using HTTP/1.1, sufficient information is typically available at the syscall level, enabling \sysname to rely on kernel-level probes for tracing. Kernel probes offer lower overhead and are generally more efficient for such use cases. 
Figure~\ref{fig:http11-mapping} illustrates an example of typical syscalls that are executed during request processing.
%\noindent\textit{Single Request Processing:} %In Figure~\ref{fig:http11-mapping}(a), we illustrate the processing flow of \textit{a single request}. 
Incoming client connections are initially placed in a queue of pending connections associated with the server's listening socket. The \texttt{accept4} system call is invoked to accept a connection from this queue, create a new socket file descriptor (fd) with the same properties as the listening socket, and return it to the calling process~\cite{accept4-ibm}. Once the connection is established and an fd is assigned, data is transferred from client to server using system calls such as \texttt{read}, \texttt{readv}, or \texttt{recvfrom} and from server to client using calls like \texttt{write}, \texttt{writev}, or \texttt{sendto}. After the server completes request processing, the fd is typically released using the \texttt{close} syscall, making it available for reuse. 
% In the case of a single request, these syscalls exhibit a distinct and sequential pattern that can be used to identify clear request boundaries. 

%Under low request rates (i.e., when handling a single request at a time), the sequence of syscalls associated with each request and its corresponding response appears in a distinct, non-overlapping order, making request boundaries easy to identify. That is, each \texttt{recvfrom} will always be associated with a \texttt{sendto}; the same goes for \texttt{accept4} and \texttt{close}, and they seldom coincide. However, when handling concurrent client requests, these patterns become interleaved and obfuscated, making accurate boundary detection significantly more complex.

%\noindent\textit{Concurrent Request Processing:} Figure~\ref{fig:http11-mapping}(b) illustrates the scenario with HTTP/1.1 pipelining, where multiple requests can be in-flight in the server. 
As previously discussed, the main challenge here is that the syscalls for multiple in-flight requests are interleaved due to concurrent worker threads, so we cannot simply pair \texttt{read/write} syscalls to identify the request begin/end. As shown in Figure~\ref{fig:http1.1-vs-http2} (left), only a single TCP connection is created per client; therefore, we can not simply use TCP connection boundaries to isolate the requests. 

%\smallskip
\paragraph{HTTP/1.1 Request \underline{Identifier}} 
Figure~\ref{code:kprobe-traces} presents a sample trace generated by eBPF probes from an HTTP/1.1 communication. The traces are captured via the bpf trace pipe~\cite{eBPF-Website}, showing selected system calls along with their arguments and return values. %eBPF enables the insertion of custom trace messages through probe instrumentation. 
Fields such as the application name, process ID, task state flags, CPU ID, and timestamp are provided by the trace pipe by default. The content following the \texttt{bpf\_trace\_printk} marker represents custom trace output generated by an eBPF probe. Recall, the main reason why we cannot easily disambiguate requests is that syscalls are associated with a process (PID), and we do not know the thread (TID) that the syscall is associated with. Thus, the key to request-disambiguation is finding metadata that is unique to individual requests.

For example, the \texttt{accept4} syscall in Figure~\ref{code:kprobe-traces} (line~6) returns file descriptor 63, and the corresponding \texttt{close} syscall (line~12) later releases that descriptor, indicating the end of that request’s connection. In contrast, the \texttt{close} in line~8 operates on file descriptor 64, which belongs to a different request. Since a file descriptor cannot be reassigned until it is released, pairing \texttt{accept4} with the matching \texttt{close} provides a reliable way to delineate request boundaries in the presence of multiple in-flight requests.

At higher request rates, the effective file-descriptor pool becomes small relative to the concurrency level, increasing descriptor reuse. \textit{We discovered that under pipelined execution, the server allocates and recycles sets of file descriptors to track in-flight requests.} Leveraging this observation, \sysname uses file descriptors as lightweight request identifiers and recovers per-request boundaries by matching \texttt{accept4} and \texttt{close} events for each descriptor, enabling accurate request disambiguation even when many requests overlap in time.

%\textit{We observe that each request sent over a connection opens up an additional file descriptor (FD) to keep track of individual requests within a connection} and is closed upon the response being sent.  %Since multiple concurrent requests can be in flight at the same time, a major challenge is that we will need to identify features to differentiate individual requests within eBPF observable events. 

% As shown in Figure~\ref{fig:http11-mapping}(c), by extracting these syscalls and their timestamp, we can obtain the latency of these requests. 

\begin{figure}[t]
\centering
\includegraphics[width=\linewidth]{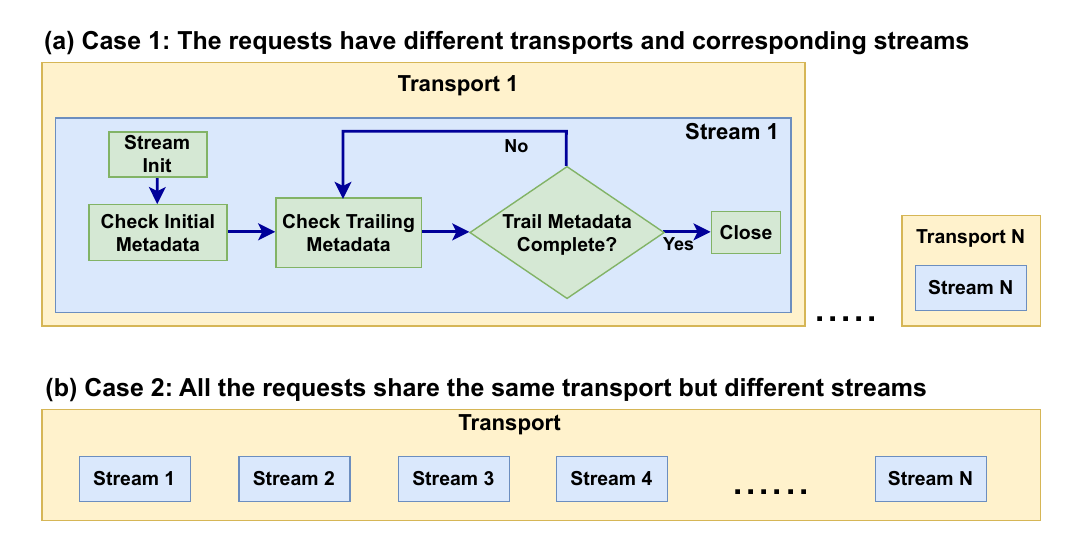}
\caption{gRPC request/stream lifecycle. (a) Some deployments create a separate transport (i.e., connection) per request, resulting in distinct transports with their own streams. (b) Other deployments multiplex multiple requests over a shared transport, using separate streams, which can become interleaved and obfuscate request boundaries.}
\vspace{-4mm}
\label{fig:gRPC-mapping}
\end{figure}

\paragraph{gRPC Request \underline{Boundary}}
gRPC builds on HTTP/2 by using binary framing and true multiplexing, allowing many requests to execute concurrently over a single long-lived connection. Similar to HTTP/1.1, connection setup and teardown are visible at the syscall layer via \texttt{accept4} and \texttt{close}. Figure~\ref{fig:gRPC-mapping} summarizes the gRPC lifecycle. However, unlike HTTP/1.1, individual gRPC requests are carried as application-level \emph{streams}: a stream represents an RPC and is not directly observable through network syscalls such as \texttt{readv/recvfrom} or \texttt{writev/sendto}. Moreover, stream activity and payload are encoded in a binary format, making per-request identification infeasible using kernel-space probes (kprobes) at the syscall level.

To recover per-request boundaries, we instead detect request arrival by probing core gRPC library functions using user-space probes (uprobes). From the gRPC library’s perspective, each RPC is associated with a distinct stream, and request boundaries can be inferred by tracking stream creation and completion. Depending on the implementation, concurrent RPCs may either allocate a \textit{separate transport} per stream (Figure~\ref{fig:gRPC-mapping}(a)) or multiplex multiple streams over a \textit{shared transport} (Figure~\ref{fig:gRPC-mapping}(b)). In the gRPC C library, streams are instantiated via \texttt{chttp2\_stream} constructor calls~\cite{gRPC_github}, followed by functions that process \textit{initial} and \textit{trailing} metadata. After the request-response exchange completes, the stream is closed, which we identify via calls to trailing-metadata completion.

\paragraph{gRPC Request \underline{Identifier}}  For different RPS rates, the streams are interleaved on the wire, making individual requests difficult to distinguish. Therefore, \sysname uses uprobes to instrument core gRPC library functions and extract \emph{transport IDs} and \emph{stream IDs} as request identifiers, allowing it to correlate start and end events and accurately recover request boundaries. As in the fd, a stream cannot be reused until the preceding request on that stream completes. \sysname identifies the start of a request by hooking the gRPC C library’s \texttt{stream\_constructor} function and identifies completion by hooking \texttt{trailing\_metadata\_completion}.

%In contrast, more typical gRPC deployments—such as that of the \texttt{Triton Inference Server}~\cite{noauthor_undated-pn} shown in Figure~\ref{fig:gRPC-mapping}(c)—employ persistent connections where multiple requests are multiplexed over individual gRPC streams. In this case, 

%While some gRPC applications still expose sufficient system call metadata to allow request tracing via kernel-level probes, others abstract away critical information, making syscall-level tracing insufficient. First, gRPC's multiplexing model reuses a single connection for all requests, encapsulating them as independent streams that may arrive out of order. Second, its use of binary encoding further obscures request metadata from the kernel, limiting visibility. To address these limitations, \sysname employs eBPF-based user-space probing to instrument key functions within the gRPC core library, enabling the extraction of request-specific metadata directly from application-level contexts. This hybrid probing strategy allows \sysname to generalize across diverse gRPC workloads and reliably collect per-request performance metrics.

\begin{table}[!b]    
\scriptsize
\centering
\vspace{-3mm}
\caption{Probes used across different workloads}
\label{tab:probe-syscall-workload}

\setlength{\tabcolsep}{5pt} % Default value: 6pt
\renewcommand{\arraystretch}{1.8}
\begin{tabular}{|c|c|c|c|c|} 
\hline
\textbf{Workload}                & \textbf{\textbf{Protocol}} & \textbf{Probe} & \makecell{\textbf{Request}\\ \textbf{Identifier}} & \makecell{\textbf{Request}\\ \textbf{Boundary}}                        \\ 
\hhline{|=====|}
\textbf{Triton}                  & HTTP1.1                   & kprobe         & accept4 fd                  & accept4 / close                                  \\ 
\hline
\textbf{Triton}                  & gRPC                       & uprobe         & \makecell{HTTP2 \\ Stream}                & \makecell{chttp2\_stream / \\ trailing\_metadata\_completion}  \\ 
\hline
\makecell{\textbf{Cloudsuite} \\ \textbf{Data caching}} & HTTP1.1                   & kprobe         & accept4 fd                  & read / sendmsg                                   \\ 
\hline
\makecell{\textbf{vSwarm}\\ \textbf{Online shop}}      & gRPC                       & kprobe         & accept4 fd                  & accept4 / close                                  \\ 
\hline
\makecell{\textbf{vSwarm} \\ \textbf{Hotel app}}        & gRPC                       & kprobe         & accept4 fd                  & accept4 / close                                  \\
\hline
\end{tabular}
\end{table}

\subsection{\sysname Probes}
Table~\ref{tab:probe-syscall-workload} summarizes the probes used by \sysname (Figure~\ref{fig:overview}~\circled{3}) to disambiguate requests across workloads.
% For each workload/protocol pair, the table reports (i) the probe type (\textit{kprobe} vs. \textit{uprobe}), (ii) the request identifier extracted by \sysname (e.g., file descriptor, transport ID, stream ID), and (iii) the events used to delineate request boundaries. 
In most HTTP/1.1 workloads, lightweight kprobes on socket-related syscalls are sufficient, and \sysname uses the file descriptor returned by \texttt{accept4} to associate the request start and end. For gRPC, \sysname selects probes based on how the application uses transports and streams (Figure~\ref{fig:gRPC-mapping}). When each stream maps to a separate socket connection, kprobes remain sufficient; when streams are multiplexed over a shared transport, \sysname relies on uprobes in the gRPC runtime to recover per-request boundaries.

% In contrast, when dealing with workloads that utilize gRPC in a more standard, stream-based fashion, individual requests do not trigger distinct syscall activity; \sysname relies on uprobes attached to key functions within the gRPC core library to identify stream-level request boundaries. 

\subsubsection{Protocol pattern variations}
Triton’s HTTP/1.1 behavior aligns with the standard pattern, and each request is associated with a connection that is opened via \texttt{accept4} and closed via \texttt{close}. We observe a different pattern in \texttt{Cloudsuite Data Caching}~\cite{palit-demystifying-cloud-benchmarking}. In this workload, connections are kept alive, and socket closures are deferred and performed in bulk at the end of execution, rather than at individual request completion. As a result, using \texttt{accept4}/\texttt{close} would overestimate per-request lifetimes. To handle this case, \sysname detects deferred/bulk closures and dynamically falls back to probing application I/O activity, using \texttt{read} as the request start signal and \texttt{sendmsg} as the response completion signal to accurately estimate request latency under persistent connections, as shown in Table~\ref{tab:probe-syscall-workload}.

gRPC exhibits two distinct usage patterns that affect what can be traced at the syscall layer. In the \texttt{vSwarm} workloads (Table~\ref{tab:probe-syscall-workload})~\cite{vswarm-benchmark}, each stream effectively results in a dedicated transport, and request processing manifests as a new socket being accepted and later closed. This \textit{one-stream-per-connection} behavior enables \sysname to use kprobes on \texttt{accept4} and \texttt{close}, with the returned file descriptor serving as the request identifier. In contrast, \texttt{Triton gRPC} multiplexes many streams over a shared transport; individual RPCs are not visible through per-request socket syscalls, and the binary framing further limits inference from packet-level events. For this optimized setting, \sysname uses uprobes on gRPC core library functions to extract transport and stream identifiers. The request boundaries are determined by matching the corresponding identifiers mentioned in Table~\ref{tab:probe-syscall-workload} from the \texttt{chttp2\_stream} construction event with \texttt{trailing\_metadata\_completion} event.

\textbf{Scope and extensibility.} \sysname is protocol-aware and does not assume that all applications follow a single \texttt{accept4/close} request pattern. Instead, it requires that each request expose a recoverable lifecycle through eBPF-observable events, either via syscalls or stable user-space hooks. We already observe such variation in our workloads: \texttt{Cloudsuite Data Caching} uses persistent connections with deferred closures, so \sysname falls back to \texttt{read/sendmsg}, while \texttt{Triton gRPC} multiplexes streams over shared transports, so \sysname uses gRPC uprobes to recover stream boundaries. Extending \sysname to other RPC or messaging frameworks, therefore, mainly requires identifying a reliable request start event, completion event, and in-flight request identifier.

% Certain applications exhibit variations that prevent straightforward request boundary detection via \texttt{accept4}. In particular, some applications, such as caching workloads~\cite{palit-demystifying-cloud-benchmarking}, do not immediately close the connections after processing a request, since the response time is minimal and persistent connections offer performance benefits. For such cases, \sysname adapts by probing alternative syscalls to accurately measure per-request latency, ensuring robustness across varied application behaviors. 

% Depending on how a workload utilizes the gRPC library, we observe that in certain cases, relying solely on kprobes and syscall activity can be sufficient to identify request boundaries, even under concurrent request processing. As illustrated in Figure~\ref{fig:gRPC-mapping}(b), the use of gRPC in the \texttt{vHive} benchmarks~\cite{vswarm-benchmark} exhibits a pattern in which each gRPC stream results in a new connection being accepted and subsequently closed. This naive usage model allows \sysname to trace request boundaries like HTTP/1.1, by attaching kprobes to the \texttt{accept} and \texttt{close} syscalls and using file descriptor IDs as the request identifiers to untangle the interleaved requests.
\subsection{How to trace with eBPF?}
%At the core of \sysname is an eBPF-based tracing framework designed to provide low-overhead instrumentation. 
Depending on the underlying communication protocol, \sysname selectively employs either kernel probes for system calls or user-space probes for gRPC library functions. 

Listing~\ref{Code:eBPFkprobe} presents an example of attaching kernel-space probes using eBPF. eBPF supports the use of kprobes for instrumenting function entry points and kretprobes for capturing return values at function exits. %In \sysnamens, syscall-specific probes are selectively attached to reduce overhead by focusing only on relevant system calls, rather than instrumenting the entire syscall interface. 
At the syscall entries, \sysname utilizes kprobes to extract essential execution context, such as the process ID (PID), syscall invocation timestamp, and relevant arguments (e.g., file descriptors and buffer pointers) for request disambiguation. This enables fine-grained filtering and request-level attribution. By combining kprobes and kretprobes at the entry and exit points of selected syscalls, \sysname captures both input parameters and return values, which leads to extracting request identifiers for request telemetry. Thus, accurately correlates low-level kernel events with high-level application behavior.

To trace gRPC library functions, \sysname employs user-space probes. \sysname attaches uprobes to user-space binaries by resolving function symbols corresponding to the target library functions. These probes extract, process, and log relevant function parameters, allowing \sysname to isolate individual requests in the same fashion as syscall-level tracing. In the case of gRPC, user-space probes enable the capture of key context information, such as transport and stream pointers, which are essential for correlating protocol-level activity with application-level behavior. %Instrumenting both kernel-space and user-space probes provides the flexibility needed for protocol-agnostic request tracing. 

\begin{figure}[!t]
% \vspace{-5mm}
\begin{lstlisting}[label=Code:eBPFkprobe, caption={Example of an eBPF kprobe (Built-in types/functions indicated in \textbf{bold}).}]
~PID_TGID~: pid and tgid of the targeted application
// Stores timestamp of accept4 return per socket fd
#BPF_HASH#(start_ts_map, #int#, #u64#);
//kprobe: triggered on entry to close syscall
[int syscall__close](#struct pt_regs# *ctx, #int# fd) {
    // Get pid_tgid of the application
    #u32# pid = #bpf_get_current_pid_tgid()# >> 32;
    if (pid != ~PID_TGID~) return 0; 
    // Temporarily store fd being closed by this pid
    fd_map.#update#(&pid, &fd);
    return 0;
}


\end{lstlisting}
\vspace{-6mm}
\end{figure}

\subsection{Request disambiguation in eBPF program}
Having established how to identify request boundaries with kprobes and uprobes, we now turn to the real-time processing and extraction of QoS metrics (Figure~\ref{fig:overview}~\circled{4}).  
To facilitate our system description, we define \textit{request-level metrics} as per-request telemetry (e.g., start time, end time, and individual request latency), and \textit{application-level metrics} as an aggregate of request-level telemetry (e.g., throughput, average latency, and tail latency). 
%HTTP/1.1 offers transparency and ease of instrumentation through system-level tools such as \texttt{strace}, the multiplexed binary nature of gRPC significantly complicates the tracking. This difference becomes critical when analyzing the performance of microservices, where gRPC is increasingly preferred for its efficiency.

%We first describe a minimal eBPF program running in the kernel that captures start and end timestamps for each request, computes per-request latency, and streams them into a ring buffer (Figure~\ref{fig:overview} \circled{4}). %Then we introduce a user-space library that continuously consumes the ring buffer and offers simple APIs for SMRs (\circled{6}).

%\subsubsection{\sysname eBPF program}
%~\newline
\subsubsection{Extracting request identifier}
Once request-level metadata is captured via probes, \sysname performs the complete request-level metrics extraction within the kernel using an eBPF program. As the application processes requests, \sysname captures start markers for individual requests through the attached probes. The request identifier (RID) is extracted from the request metadata. RIDs are used to correlate start and end events. They may vary depending on the probe type, typically a file descriptor for kprobes or a stream ID for uprobes, as summarized in Table~\ref{tab:probe-syscall-workload}. For instance, in most benchmarks, a successful \texttt{accept4} syscall returns a file descriptor, which is later passed as an argument to the corresponding \texttt{close} syscall. Similarly, in the \texttt{CloudSuite Data Caching} server, although \texttt{read} and \texttt{sendmsg} are used to track request activity, they operate on file descriptors initially assigned by \texttt{accept4}. 

\subsubsection{Tracking request latency}
After extracting a unique request identifier (RID) for each request, \sysname stores it in an \textit{eBPF hash map} along with the corresponding start timestamp. This map is implemented using \texttt{BPF\_HASH}, a key-value data structure provided by eBPF~\cite{eBPF-Website}. When the probe associated with the request's completion is triggered, \sysname retrieves the RID, matches it against the stored entry, and computes the request’s latency. The resulting per-request metrics are then logged for consumption by the user-space library to build up application-level metrics.

\subsection{Efficient data streaming to userspace library}
Rather than computing application-level metrics (e.g., tail latency, throughput) within the eBPF program, \sysname captures a minimal set of request-level information in the eBPF program and streams it to user space using an eBPF ring buffer~\cite{eBPF-Website} (Figure~\ref{fig:overview}~\circled{5}). The eBPF ring buffer is a kernel-to-user-space communication mechanism optimized for efficiently streaming structured data with minimal overhead. This design offloads computation from the kernel, preserves low runtime overhead, and circumvents eBPF limitations such as the absence of floating-point support, while enabling more flexible and expressive analysis in user space. 

Each request is captured in a structured format containing a start timestamp and computed latency, which is pushed to the ring buffer in shared memory. 
% This processing occurs in parallel with normal application execution, ensuring no disruption or observable performance degradation. 
Crucially, \sysname relies solely on precise start and end markers, without requiring feedback loops or heuristics, to deliver accurate, real-time request-level metrics.

%\DANIEL{Should we discuss what eBPF shared memory is? Maybe discuss that eBPF comes with pre-defined data structures. It's not clear if we implemented these data structures or if they're already provided. }\MUNTAKA{Both the hash map and ring buffer are eBPF-provided data structures, mentioned them, and also cited}

%\noindent\textbf{Other considerations.}
%\DANIEL{What is the size of the ring buffer? How do we size it? What if the size is too small?}\MUNTAKA{Buffer size is described in the next section}

\subsection{\sysname Userspace Library} 
\sysname includes a lightweight user-space library that (a) initializes eBPF probes and shared memory buffers, and (b) builds application-level metrics from the per-request metadata streamed from the eBPF program. %This unified library also manages deployment, data ingestion, low-overhead statistics computation, and telemetry delivery to client runtimes that abstract away all eBPF-specific details from the SMRs.

% % \noindent\textbf{Library initialization: }
% During initialization, the \sysname user-space library first verifies kernel capabilities, then observes traces from the target application to determine the appropriate probe types to deploy. Based on this assessment, it loads the corresponding eBPF object and allocates a per-CPU kernel ring buffer. \sysname supports dynamic instrumentation via the \texttt{probe\_add(pid\_t, type, loc)} and \texttt{probe\_remove(pid\_t, type, loc)} APIs, where \texttt{pid\_t} specifies the target process, \texttt{type} indicates the probe kind (kprobe or uprobe), and \texttt{loc} specifies the attachment point (i.e., syscall for kprobes or user-space function for uprobes). These interfaces enable SMRs to enrich the tracing context on demand. Additionally, SMRs can control application-level tracing dynamically using \texttt{enable\_tracing(pid\_t)} and \texttt{disable\_tracing(pid\_t)}. These functionalities let the SMRs start or end the application tracing without requiring restarts.
% %(minimum 6.1 LTS with \texttt{BPF\_RINGBUF\_MAP\_TYPE}), 
% During initialization, \sysname allocates a ring buffer with a default size of 64 kB (8 pages), which can be tuned by the caller to accommodate different workload intensities. This buffer is directly accessible from user space via the \sysname library.

\begin{table}[t]
\footnotesize
\centering
\caption{Supported application-level metrics and the four high-demand API calls corresponding to the metrics. }
%\DANIEL{What are API calls to initialize the library (buffers) and setup the probes? }\MUNTAKA{These are the four high demand API calls that the SMRs need most. They are always calculated and ready for the SMRs}
\begin{tabular}{|p{3cm}|p{4cm}|}
\hline
\textbf{Metric} & \textbf{API Call} \\
\hline
Current requests per second & \texttt{get\_RPS(pid)} \\
\hline
Latency of the most recent request & \texttt{get\_latest\_latency(pid)} \\
\hline
Moving-average latency & \texttt{get\_average\_latency(pid)} \\
\hline
99th-percentile latency (sliding window) & \texttt{get\_latency\_percentile (pid, p)} \\
\hline
\end{tabular}
\label{tab:metrics-api}
%\vspace{-5mm}
\end{table}

\subsubsection{Initializing the Library}
On startup, \sysname library first checks for the required eBPF kernel features. It then provides \texttt{start\_tracing(pid\_t)} and \texttt{stop\_tracing(pid\_t)} APIs to the SMRs. When \texttt{start\_tracing} is called, the library launches probes for all candidate events and monitors for probes with activity. It prunes any idle probes and retains only those that capture useful data. It also allocates ring buffers to access information gathered by the probes. Once this setup is complete, the eBPF component begins streaming per-request latency records to the user-space library. By default, each ring buffer is 128 kB in size. In all of our experiments, this capacity was sufficient for our highest-throughput benchmark (no buffer overflows), and we saw no benefit from larger buffers. If needed, users can adjust the buffer size via the library API.

The \sysname library also maintains per-benchmark ring buffers hosted in user space, which store the latest latencies to decouple low-level per-request metadata collection from application-level metrics computation. By default, this buffer stores the last 100k latencies, which is sized to support high-throughput applications like memcached without incurring performance penalties. Moreover, this enables \sysname to be able to handle multi-workload scenarios, making it more generalizable and broadly applicable for use in diverse and dynamic SMR environments.

\subsubsection{\sysname APIs and maintaining application-level metrics}
The \sysname library provides SMRs with continuous, low-overhead access to latency and throughput metrics through four frequently-used APIs (Table~\ref{tab:metrics-api}). Instead of scanning the entire ring buffer each time SMRs request data, the library periodically maintains aggregated statistics.

Specifically, request throughput (\texttt{get\_RPS(pid)}) is computed by dividing the total number of requests currently stored by the elapsed time between the oldest and newest entries in the buffer.  This calculation occurs in constant time, $O(1)$, since timestamps are directly accessible. Similarly, the most recent request's latency value (\texttt{get\_latest\_latency(pid)}) is immediately returned, also in $O(1)$.

To efficiently compute the moving-average latency (\texttt{get\_average\_latency(pid)}), the library maintains a running sum of latencies. Each time a new latency value arrives, the library adds it to the sum and subtracts the oldest value, keeping the calculation in $O(1)$. For tracking tail latencies (\texttt{get\_latency\_percentile(pid, p)}), such as the 99th percentile, the library maintains two balanced multisets. These data structures enable incremental updates to the percentile calculation in $O(\log n)$ with each new request. It is worth mentioning that this approach is only used to speed up the API call for the most used statistic in our experiments, which is the 99th percentile. For other tail latency requests, the complete ring buffer needs to be scanned.

% \MOE{I think this is implied to this point}
% \subsubsection{Support for Multiple Applications}

\section{Evaluation}
\label{section:Evaluation}
\begin{figure*}[t]
\centering
\includegraphics[width=\textwidth]{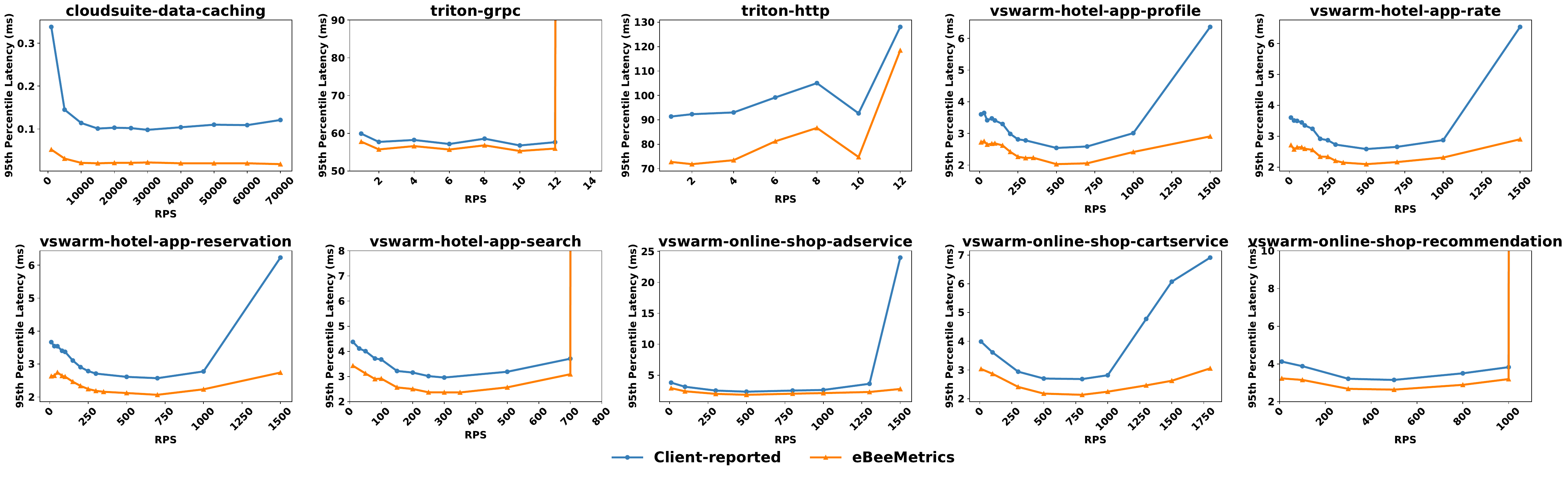}
\caption{Comparison of workload's measured tail latency as reported by client and by \sysnamens. \sysname closely tracks the measured latency for each RPS with high accuracy in all ten benchmarks. Measured gap is due to network latencies.}
\label{fig:client-vs-server-latency}
\end{figure*}

\begin{figure*}[t]
\centering
\includegraphics[width=\textwidth]{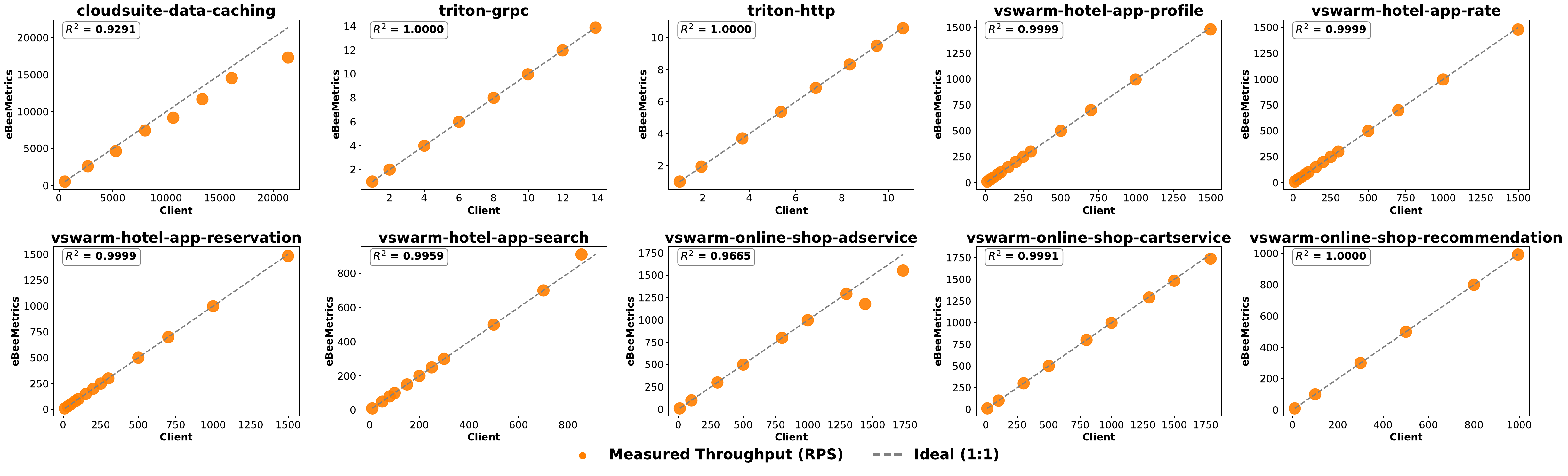}
\caption{Comparison of the workload's throughput as reported by the client and by \sysnamens. \sysname is capable of matching the measured throughput for each RPS with high accuracy in all ten benchmarks.}
\label{fig:client-vs-server-throughput}
\vspace{-3mm}
\end{figure*}

We now evaluate the efficacy of \sysname estimates of request-level and application-level metrics. 

\noindent\textbf{Server configuration: }
We evaluated \sysname on a high-end server with a dual socket 16-core AMD EPYC 7302, 512GB of memory, and running Ubuntu 20.04 with Linux 5.15.0-89. The \sysname library is agnostic to processor types and only requires a Linux kernel that supports eBPF.%\footnote{eBPF support was first introduced in Linux 3.15 in June 2014}. %\DANIEL{When did eBPF begin being supported?}\MUNTAKA{eBPF was introduced and began receiving official support in the Linux kernel version 3.15, which was released in June 2014.}

\noindent\textbf{Workloads: }
We evaluate \sysname with latency-critical applications from two benchmark suites and a state-of-the-art ML inference server. This includes seven different micro-benchmarks from the 33 standalone benchmarks of \texttt{vSwarm}~\cite{vswarm-benchmark}, specifically \texttt{Hotel Reservation}~\cite{10.1145/3297858.3304013} and \texttt{Online Shop}~\cite{google-online-boutique}. We chose \texttt{search}, \texttt{reservation}, \texttt{rate}, and \texttt{profile} from \texttt{Hotel Reservation} and \texttt{recommendation}, \texttt{adservice}, and \texttt{cartservice} from \texttt{Online Shop}. 
%These are chosen based on the high dependency and the degree of interconnection between the services to increase the generalizability of \sysnamens. \DANIEL{What does this last sentence mean?} 
Note, we evaluated \sysname with other micro-benchmarks from the \texttt{vSwarm} suite and observed they all behaved similarly from a gRPC point of view; thus, for brevity, we only selected these subsets as representative of the \texttt{vSwarm} suite. 

Furthermore, we evaluated the latency-critical application \texttt{Data Caching (Memcached)} from the \texttt{CloudSuite} benchmarks suite~\cite{palit-demystifying-cloud-benchmarking} and \texttt{Triton Inference Server}, an open-source deep-learning inference server developed by NVIDIA \cite{noauthor_undated-pn, noauthor_2020-vl}, which supports both HTTP and gRPC protocols as its inference API, providing us with a robust comparison against different protocols. 

%All workloads are executed within Docker containers, with separate client and server machines, to capture both server and network behavior. 
All workloads handle concurrent requests and exhibit a wide range of request-handling software threading behavior. This demonstrate that \sysname is agnostic to an application's threading and queuing design. For example, \texttt{Data Caching} provides a straightforward request-handling threading behavior where each thread consumes and processes a request. \texttt{Triton} has dedicated threads that consume requests and dispatch them across separate threads for processing.% The \texttt{vSwarm} benchmark focuses on realistic data-intensive workloads.
%\DANIEL{Is saying vSwarm benchmarks are data-intensive correct? }

\subsection{\sysname Metrics}

\noindent\textbf{Latency: }
In Figure \ref{fig:client-vs-server-latency}, we sweep a range of RPS for each workload and plot the 95th percentile tail latency for two scenarios. \textit{Client-reported} plots the tail latency as reported by the client generating the requests. This is the default tail latency reported by the workloads. The \textit{eBeeMetrics} scenario plots the tail latency reported by \sysnamens. Overall, \textit{\sysname} accurately tracks the \textit{Client-reported} latency.

Since \sysname tracks request boundaries when it enters and exits the server, the tail latency reported is entirely \textit{server-side latency} (including server-side queueing time and server-side processing time), whereas the \textit{Client-reported} scenario also includes network latency, which is shown as the fairly consistent gap between both scenarios. %  see that \sysname can accurately measure the average latency for each benchmark until the client reaches the threshold. 
%The gap between the client and server is due to the network latency. Client-reported latency includes network latency, server-side queuing time, and server-side processing time. Since \sysname only measures server-side latency, the request latency only consists of server-side queuing time and processing time. 
%For all workloads, the gap between blue and green is fairly consistent, reflecting the constant network latency. 
\texttt{vSwarm} workloads tend to see a 0.75 ms network latency, while \texttt{Triton} observes 20 ms and 3 ms for HTTP and gRPC protocols, respectively.  However, some start diverging, such as \texttt{vSwarm Online Shop Cart} or \texttt{Online Shop Ad}, where the network becomes the bottleneck with network congestion since server-side request time is still fairly consistent. \textit{This demonstrates that \sysname can be used to supplement existing client-reported tail latency to help identify whether tail latency increases are due to network-side or server-side dynamics. }

The biggest difference between client-reported and \sysname latency occurs with \texttt{Memcached}, which has the highest throughput and shortest request processing time. Most requests only spend about 0.010 ms in the server, with client-reported latency of 0.1 ms, indicating the majority of observed latency is due to networking. This is because \texttt{Memcached} is a network-bound workload with simple request processing. 
%\DANIEL{Since memcached never saturates the CPU since it is network bound, shouldn't we still see a spike in latency when the network is saturated? }\MUNTAKA{The spike didn't occur even at 70k rps, but throughput plot shows it started diverging from 20k rps}

\begin{figure*}[!t]
\centering
\includegraphics[width=0.9\textwidth]{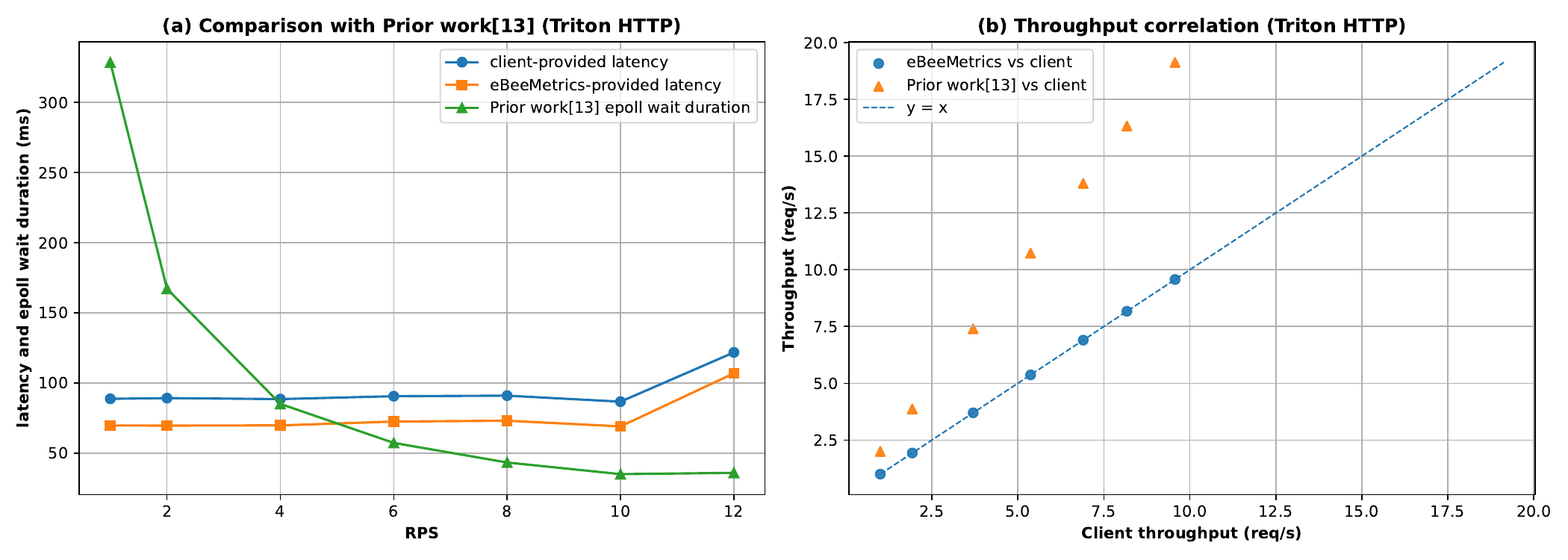}
\caption{Comparison with prior work~\cite{Paper:kernelobservability} on Triton HTTP. (a) \sysname latency tracks client-reported latency across RPS, while epoll\_wait from~\cite{Paper:kernelobservability} shows a different trend because it is a slack proxy, not a latency metric. (b) \sysname throughput closely matches client throughput, whereas the proxy from~\cite{Paper:kernelobservability} increasingly deviates at higher request rates.}
\label{fig:compare}
%\vspace{-3mm}
\end{figure*}

\begin{figure*}[!t]
\centering
\includegraphics[width=0.95\textwidth]{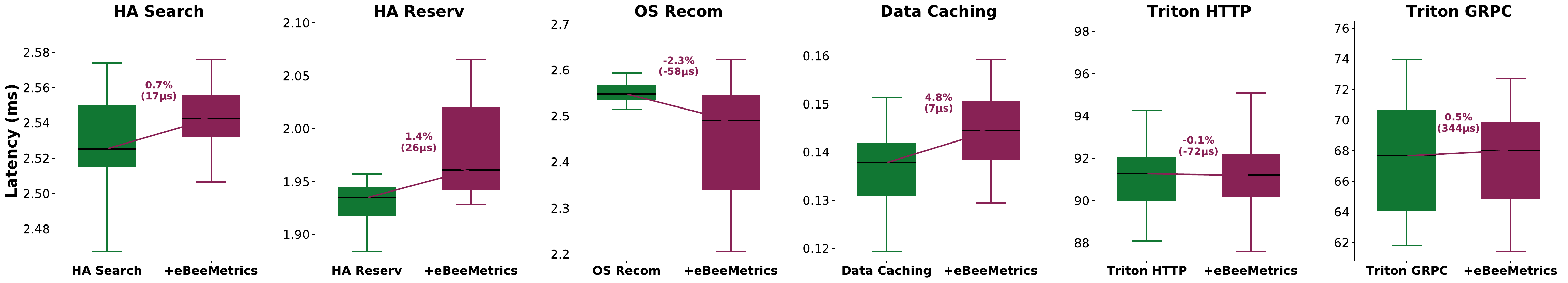}
\caption{Latency distribution w/ and w/o \sysnamens. Arrows indicate the percentage overhead introduced by \sysnamens.}
\label{fig:overhead}
\vspace{-3mm}
\end{figure*}

\noindent\textbf{Throughput: }
Figure~\ref{fig:client-vs-server-throughput} shows the client-reported throughput vs \sysnamens-reported throughput. For the majority of cases, the throughput matches \textit{exactly} with most outliers occurring at the higher end of the RPS. As shown previously in Figure~\ref{fig:client-vs-server-latency}, these outliers are a result of network saturation. 

% \vspace{-1mm}
\subsection{Comparing \sysname with Prior Work~\cite{Paper:kernelobservability}}
Prior work~\cite{Paper:kernelobservability} uses eBPF-derived syscall traces to infer offline throughput-oriented proxy signals. In contrast, \sysname reconstructs both throughput and per-request latency online from streaming eBPF events. Figure~\ref{fig:compare} compares the two on the \texttt{Triton HTTP} workload. In Figure~\ref{fig:compare}(a), \sysname latency closely tracks client-reported latency across RPS, while the average epoll\_wait duration from~\cite{Paper:kernelobservability} follows a different trend because it reflects event-loop slack rather than per-request service time. The small gap between client and \sysname latency is consistent with client measurements, including network delay. In Figure~\ref{fig:compare}(b), \sysname throughput coincides with client throughput and remains on the ideal \textit{y=x} line, while the metric from~\cite{Paper:kernelobservability} deviates at higher request rates, consistent with its role as an offline proxy rather than a direct request-count measurement. Overall,~\cite{Paper:kernelobservability} is useful for offline slack and saturation analysis, whereas \sysname is better suited for online runtimes requiring accurate throughput and request-level latency feedback.

% \vspace{-1mm}
\subsection{\sysname Overhead}
Figure~\ref{fig:overhead} illustrates the latency distribution across various benchmarks, both when executed standalone and when instrumented with \sysname tracing. \sysname requires no more than three probes per application, keeping instrumentation overhead low. The \texttt{Triton} servers exhibited negligible overhead, while the \texttt{vSwarm} servers experienced no more than a 3\% increase. The figure highlights the three most latency-sensitive workloads among the seven \texttt{vSwarm} benchmarks. The highest overhead observed in the \texttt{CloudSuite Data Caching(Memcached)} server is attributable to its inherently low-latency operations. Even so, the additional latency is minimal, averaging around 8$\mu$s. The negative overheads are due to run-to-run variation, indicating that \sysnamens's overheads are within the margin of measurement error.

\vspace{-3mm}
\subsection
{Integrating \sysnamesec~with System Management \\ Runtimes}
\label{section:Case Study}

\begin{figure}[t!]
\includegraphics[width=\linewidth]{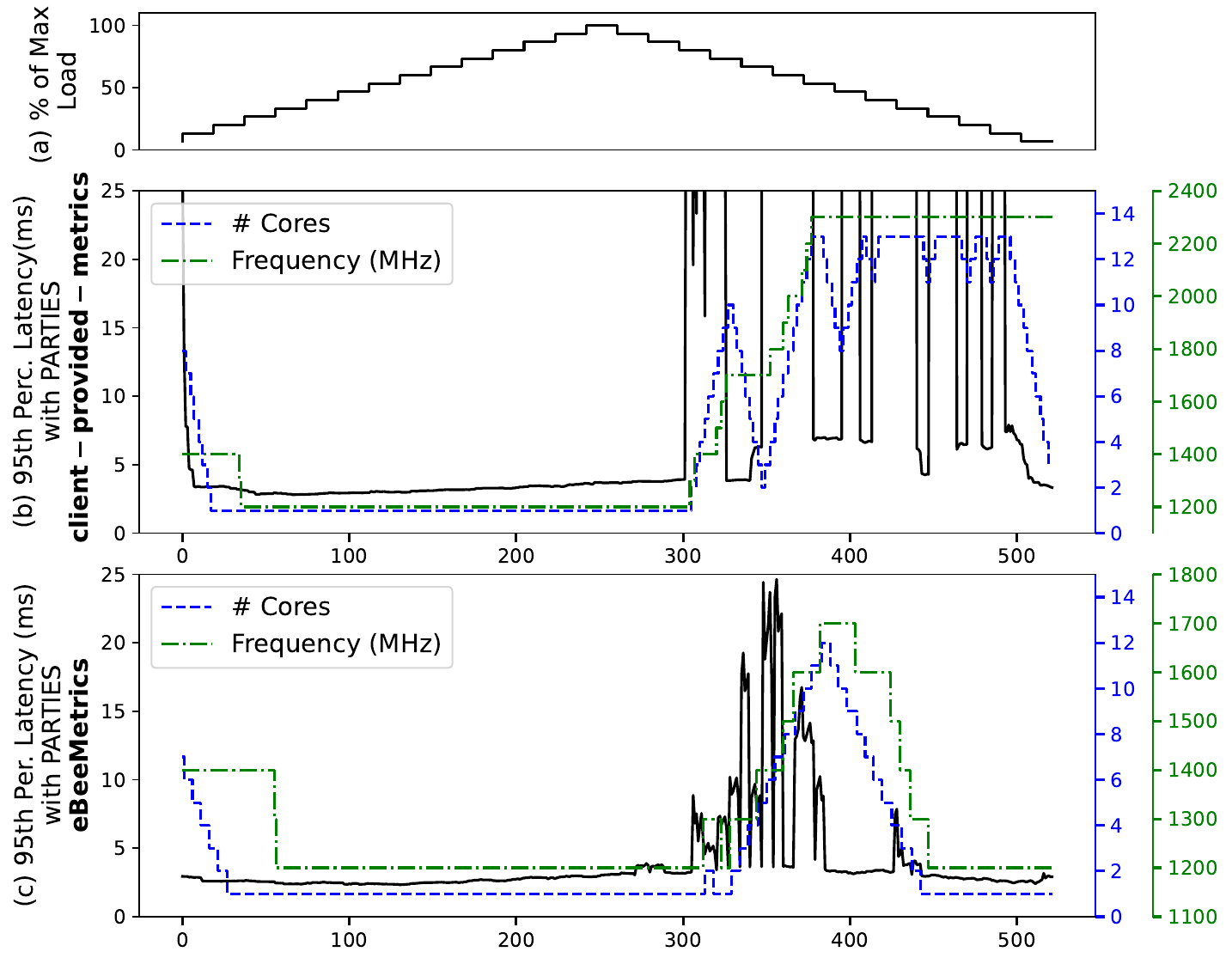}
\vspace{-4mm}
\caption{PARTIES resource manager ~\cite{10.1145/3297858.3304005} w/ and w/o \sysnamens under load pattern (a). 
(b) illustrates latency and resources controlled by PARTIES with application-provided metrics and (c) with feedback metrics provided by \sysnamens.}
% Both scenarios lead to acceptable latency levels (less than 25ms).
% \DANIEL{Can you make both y-axis go up to 25ms? Also, I took a closer look. I thought latency was frequency. Both cores and frequency seem to be static, even when the load changes. I would expect freq/core to adjust as load increases/decreases...}\MUNTAKA{Updated}
\label{fig:Parties}
\vspace{-4mm}
\end{figure}

To demonstrate a use-case of integrating \sysname with a system management runtime, we utilize PARTIES~\cite{10.1145/3297858.3304005}, which manages resources of co-located latency-critical and best-effort workloads without violating QoS requirements. QoS is maintained by computing the latency slack available for each request, and assigning co-located applications different amounts of Last Level Cache (LLC) ways, CPU cores, and varying CPU frequency to achieve minimum power consumption without violating QoS requirements. PARTIES increase resource allocation to applications near the QoS target or with QoS violation, and will reduce the resources allocated to applications that are well below the QoS target.

In PARTIES, the application clients directly report the per-request latency metrics to PARTIES. 
In this case study, we present a plug-in replacement of PARTIES' reliance on client-side tail latency information with \sysnamens' metrics. %This integration required no modifications to PARTIES or application instrumentation, as \sysname relies solely on metrics extracted via eBPF. 
%PARTIES use request latency feedback to calculate the amount of available latency slack. 
Therefore, we simply modify the PARTIES code to read the latencies extracted from \sysname instead of its original client-provided latency.

\subsubsection{Experiment Setup} 
PARTIES relies on Intel’s Cache Allocation Technology (CAT)~\cite{noauthor_undated-en, Guide_undated-nl} to manage LLC way partitioning. However, our server uses an AMD processor, which does not support CAT. As a result, we focused on managing CPU cores and frequency settings.

We selected two latency-sensitive workloads for co-location: \texttt{Hotel App Search} from the vSwarm benchmark suite and \texttt{Data Caching (Memcached)} from CloudSuite. These applications serve as representative case-study workloads for our integration.

Following the PARTIES methodology, our experiments maintained a constant, low-intensity load on the \texttt{Data Caching} application (below 20\% of its maximum capacity) to ensure stable resource allocation. By contrast, \texttt{Hotel App Search} was ramped from 7\% load to 100\% saturation and then ramped back down to its initial low-load level. We set a QoS latency target of 25ms. This dynamic load pattern, along with PARTIES’ corresponding resource allocation behavior, is depicted in Figure~\ref{fig:Parties}. 

%For the original PARTIES setup shown in Figure~\ref{fig:Parties}(b), we specified a QoS latency target of 25 ms. However, since \sysname captures latency from the server side, we set a slightly lower QoS target of 24 ms for the configuration in Figure~\ref{fig:Parties}(c) to exclude network delays and ensure a fairer comparison. 

\subsubsection{Results}
For both default (b) and \sysname (c) in Figure~\ref{fig:Parties}, when application latency approached the threshold, PARTIES increased resource allocation. Whereas, when latency remains below the target, it scales back resources to conserve power. 

As shown in Figure~\ref{fig:Parties}(b), PARTIES initially reduces the number of allocated cores and lowers CPU frequency, since application latency remains well below the 25ms QoS target. As the load on \texttt{Hotel App Search} increases and maintaining the latency target becomes more challenging, PARTIES responds by allocating additional resources. If latency exceeds the 25-ms threshold, further resources are provisioned to bring performance within the desired bounds. During the load decrease phase, resource allocations are scaled down accordingly.

Figure~\ref{fig:Parties}(c) shows the system's behavior when \sysname is used in place of application-side instrumentation, without relying on explicit QoS targets or client-provided latency inputs. The overall trend mirrors that of Figure~\ref{fig:Parties}(b). However, key differences emerge due to \sysname only capturing server-side latency, which excludes the impact of network queuing. While both scenarios observe spikes in tail latency around the same time and increase resource allocation, \sysname detects reduced latency sooner and scales down resource usage more aggressively while the baseline is still observing network queueing effects. 

This demonstrates that \sysname can be beneficial in supplementing other sources of QoS measurements to identify server-side and network-side sources of tail latency. Since resource allocation of core frequency and core allocation impacts server-side processing and not network-side queueing, QoS feedback signals from server-side only tail latency (\sysnamens) may be more beneficial than client-provided. %Therefore, 

\section{Conclusion}
\label{section:Conclusion}
This paper presents \sysnamens, a lightweight eBPF-based library that enables accurate, real-time collection of QoS feedback metrics, without requiring application instrumentation or direct client feedback. \sysname captures key request telemetry with minimal overhead and operates entirely from the server side. It currently supports HTTP/1.1 and gRPC/HTTP/2, and its design can be extended to other RPC or messaging frameworks that expose stable request start/end hooks or reliable syscall-visible request lifecycles. The \sysname tool is open-source; the source code is available at: \url{https://github.com/Ibnathism/eBeeMetrics}.
% It supports latency-sensitive applications running over modern protocols such as HTTP and gRPC, and scales to high-throughput workloads without interfering with application behavior. We plan to open-source the \sysname tool when the paper is accepted and published.

\section*{Acknowledgements}
We would like to thank the anonymous reviewers for their invaluable comments and suggestions. This work is partly supported by the University of California, Riverside, the National Science Foundation under grants CNS-1955650, CNS-2047521, CCF-2324940, and CCF-2324941, and the eBPF Foundation Academic Research Grant Program. In addition, artifact evaluation was provided using the Chameleon testbed~\cite{keahey2020lessons} supported by the National Science Foundation.
%%%%%%% -- PAPER CONTENT ENDS -- %%%%%%%%

%%%%%%%%% -- BIB STYLE AND FILE -- %%%%%%%%
\bibliographystyle{IEEEtran}
\bibliography{refs}
%%%%%%%%%%%%%%%%%%%%%%%%%%%%%%%%%%%%

\appendix
\section{Artifact Appendix}

%%%%%%%%%%%%%%%%%%%%%%%%%%%%%%%%%%%%%%%%%%%%%%%%%%%%%%%%%%%%%%%%%%%%%
\subsection{Abstract}

This artifact contains the full implementation of \sysnamens, an eBPF-based library for online observability of QoS metrics (throughput and latency) from server-side events. The artifact includes: (1) eBPF kernel programs implementing kprobes and uprobes, (2) A userspace library exposing APIs for retrieving QoS metrics, and (3) Scripts and documentation to reproduce latency and throughput evaluation experiments.

Using this artifact, the evaluators were able to reproduce the main result of the paper: accurate server-side measurement of request latency (Figure~\ref{fig:client-vs-server-latency}) and throughput (Figure~\ref{fig:client-vs-server-throughput}) without application instrumentation. For artifact evaluation, a representative subset of the workloads used in the paper is provided, focusing on the Triton Inference Server benchmark.
% Using this artifact, reviewers can reproduce the main result of the paper: accurate server-side measurement of request latency (Figure~\ref{fig:client-vs-server-latency}) and throughput (Figure ~\ref{fig:client-vs-server-throughput}) without application instrumentation. For ease of artifact evaluation, we provide a subset of the workloads used in the paper, focusing on the Triton inference server benchmark.

\subsection{Artifact Check-list (Meta-information)}

{\small
\begin{itemize}
  \item {\bf Algorithm: } Feedback-free online request disambiguation from syscall traces and user-space functions
  \item {\bf Program: } eBPF programs (C) and userspace library (Python)
  \item {\bf Compilation: } clang/LLVM; gcc/g++
  \item {\bf Binary: } eBPF object files and userspace executable
  \item {\bf Data set: } Triton Inference Server benchmark
  \item {\bf Run-time environment: } Linux with eBPF support
  \item {\bf Hardware: } x86-64 server
  \item {\bf Execution: } Server-side tracing with client workload generation
  \item {\bf Metrics: } Request throughput (RPS), latest latency, average latency, percentile latency
  \item {\bf Output: } Server-side latency and throughput measurements, per-request latency measurements
  \item {\bf Experiments: } Accuracy comparison and one-to-one correlation with the client-provided metrics
  \item {\bf How much disk space is required (approximately)?: } ~20 GB (mainly for the Triton Docker image)
  \item {\bf How much time is needed to prepare workflow (approximately)?: } ~1-2 hours
  \item {\bf How much time is needed to complete experiments (approximately)?: } ~30-45 mins
  \item {\bf Publicly available?: } Yes
  \item {\bf Code licenses (if publicly available)?: } MIT
  \item {\bf Workflow automation framework used?: } Yes
  \item {\bf Archived (provide DOI)?: } Yes
\end{itemize}
}

%%%%%%%%%%%%%%%%%%%%%%%%%%%%%%%%%%%%%%%%%%%%%%%%%%%%%%%%%%%%%%%%%%%%%
\subsection{Description}

\subsubsection{How to access}
\sysname source code, installation walkthrough, and experiment instructions are publicly available at the Github repository (https://github.com/Ibnathism/eBeeMetrics) and Zenodo (https://doi.org/10.5281/zenodo.18895625). 

% For ease of access, we provided a Chameleon Cloud ssh access with everything pre-set up for experimentation. 

\subsubsection{Hardware dependencies}
The artifact was tested on a local server with dual-socket AMD EPYC 7302 processors and a Chameleon Cloud baremetal instance with dual-socket Intel Xeon Gold 6240R processors.

\subsubsection{Software dependencies}
\begin{itemize}
    \item Linux kernel 5.8 or newer
    \item eBPF support enabled in the Linux kernel
    \item Docker
    \item Root privileges for running the eBPF tracing
\end{itemize}

%%%%%%%%%%%%%%%%%%%%%%%%%%%%%%%%%%%%%%%%%%%%%%%%%%%%%%%%%%%%%%%%%%%%%
\subsection{Installation}
% All dependencies and workloads are already installed and configured on a Chameleon Cloud instance. Reviewers only need to do the following:
% \begin{itemize}
%     \item Log into the Chameleon instance: \begin{verbatim} ssh test@129.114.109.252 \end{verbatim} (Password: \texttt{ubuntu}). 
%     % The user \texttt{test} has sudo privileges.
%     \item Navigate to the repository: \begin{verbatim} cd eBeeMetrics \end{verbatim}
%     \item Run the experiments and generate plots by following the instructions in \texttt{docs/Latency-and-throughput-plots.md} from the repository
% \end{itemize}

% If you prefer to reproduce the artifact without the preconfigured Chameleon environment, 
The full setup can be reproduced on any Linux machine with eBPF support. For detailed instructions, clone the repository (https://github.com/Ibnathism/eBeeMetrics), refer to the \texttt{README} \textit{(Section: Running Everything From Scratch)}.

%%%%%%%%%%%%%%%%%%%%%%%%%%%%%%%%%%%%%%%%%%%%%%%%%%%%%%%%%%%%%%%%%%%%%
\subsection{Experiment workflow}
To run the experiments and generate plots, follow the instructions in \texttt{docs/Latency-and-throughput-plots.md} in the repository. The experiment workflow follows the steps below:
\begin{enumerate}
% \item Log into the provided Chameleon Cloud instance and navigate to the artifact repository.
\item Ensure the Triton server and client containers are running using \texttt{docker ps}.
\item Setup and start the \sysname tracing library. During execution, the library reports QoS metrics (RPS and latency statistics) via its API in real time.
\item Launch the Triton client workload to generate inference requests.
\item After the experiment completes, the collected server-side latency data is stored in the directory: \texttt{lib/latencies/}.
\item Copy the client-reported latency file from the client container.
\item Activate the virtual environment and run the plotting scripts in the \texttt{latencies} directory to generate latency and throughput plots.
\end{enumerate}

%%%%%%%%%%%%%%%%%%%%%%%%%%%%%%%%%%%%%%%%%%%%%%%%%%%%%%%%%%%%%%%%%%%%%
\subsection{Evaluation and expected results}

Using the provided scripts, a subset of the results presented in Figures~\ref{fig:client-vs-server-latency} and~\ref{fig:client-vs-server-throughput} of the paper can be reproduced. The generated plots show:
\begin{itemize}
    \item Comparison of \sysnamens-reported tail latency and client-reported tail latency
    \item Throughput comparison between client measurements and \sysname
\end{itemize}
The plots should demonstrate a strong correlation between server-side metrics collected by \sysname and the client-provided metrics. 
% Since the Chameleon cloud instance used for artifact evaluation is a different machine from the one used for the paper's original experiments, 
Since the results depend on the environment used for the experiment, the specific results will differ, though the overall trends should remain consistent. The expectation is that Figure~\ref{fig:client-vs-server-latency} should have both client-reported and \sysname lines track each other closely, while Figure~\ref{fig:client-vs-server-throughput} should have points lie mainly on the ideal line with $R^{2}$ value close to 1. 
% Figure~\ref{fig:overhead} is not included in the artifact because overhead measurements depend strongly on hardware configuration and system load. Figure~\ref{fig:Parties} is also not included since it requires integration with the PARTIES resource manager and specific hardware support.

%%%%%%%%%%%%%%%%%%%%%%%%%%%%%%%%%%%%%%%%%%%%%%%%%%%%%%%%%%%%%%%%%%%%%
\subsection{Experiment customization}
Users may modify the experiments by adjusting request rates, duration, or the percentile latency reported by the library. Additional workloads can also be traced by attaching the \sysname probes to other server applications. The \texttt{lib/lib.py} includes functionality for all the workloads experimented with in this paper.

%%%%%%%%%%%%%%%%%%%%%%%%%%%%%%%%%%%%%%%%%%%%%%%%%%%%%%%%%%%%%%%%%%%%%
\subsection{Notes}
Root privileges are required to load eBPF programs. The Linux kernel must support eBPF and BPF JIT compilation. 

%%%%%%%%%%%%%%%%%%%%%%%%%%%%%%%%%%%%%%%%%%%%%%%%%%%%%%%%%%%%%%%%%%%%%
\subsection{Methodology}

Submission, reviewing, and badging methodology:
\begin{itemize}
  \item \url{https://www.acm.org/publications/policies/artifact-review-and-badging-current}
  \item \url{https://cTuning.org/ae}
\end{itemize}

\end{document}